\documentclass[prl,twocolumn,showpacs,showkeys,amssymb]{revtex4}
\usepackage{graphicx}

\begin{document}

\title{Passive Sliders on Fluctuating Surfaces: Strong-Clustering States}
\author {Apoorva Nagar $^{1}$, Mustansir Barma $^{1}$ and Satya N. Majumdar $^{2}$}
\address{$^1$ Department of Theoretical Physics, Tata Institute of Fundamental Research, Homi Bhabha Road, Mumbai 400 005, India \\
$^2$ Laboratoire de Physique Theorique et Modeles Statistiques,\\ 
Universit'e Paris - Sud, B$\hat{a}$t 100, Orsay, France \\
}

\date{\today}
\begin{abstract}
We study the clustering properties of particles sliding downwards on a fluctuating surface evolving through the Kardar-Parisi-Zhang equation, a problem equivalent to passive scalars driven by a Burgers fluid. Monte Carlo simulations on a discrete version of the problem in one dimension reveal that particles cluster very strongly: the two point density correlation function scales with the system size with a scaling function which diverges at small argument. Analytic results are obtained for the Sinai problem of random walkers in a quenched random landscape. This equilibrium system too has a singular scaling function which agrees remarkably with that for advected particles.
\end{abstract}


\maketitle
  
Interesting correlations develop in a system of particles which are driven by a fluctuating field but do not react back on it. This problem of semiautonomously coupled systems is of general interest and importance. In the context of fluid dynamics, it comes up as the passive scalar problem, where a density field (e.g. dye particles) or a temperature field is advected by a stirred fluid ~\cite{shraiman,falkovich1}. The interplay of fluid advection and random thermal noise can lead to two distinct behaviours of the passive scalar field --- it may either spread out ~\cite{shraiman,kraichnan}, or show a tendency to collapse and form clustered aggregates. The clustered state can arise when either the passive scalar
flow ~\cite{falkovich} or the fluid itself ~\cite{gawedzki} is compressible. Clustering also arises in models of thin film growth in binary systems where domain walls between the two different species can act like passive particles while surface growth is the driving field  ~\cite{drossel1,drossel2}.\\

In this paper, we study the clustered state that occurs in a simple but nontrivial model, namely non-interacting, passive particles sliding downwards on a fluctuating one-dimensional interface described by the Kardar-Parisi-Zhang (KPZ) equation. This problem can be mapped onto that of particles advected by a noisy Burgers fluid. Some aspects of this problem have been earlier studied by Drossel and Kardar~\cite{drossel1,drossel2}. However, the questions we study here are different and our results indicate that the system reaches a new type of steady state, namely the strong-clustering state (SCS). The signature of the SCS is the divergence at origin of the two-point density-density correlation function as function of the separation scaled by the system size. This divergence provides a measure of clustering in the system, and distinguishes the SCS from normal phase separated states~\cite{bray} and also from fluctuation-dominated phase separated states~\cite{das,das1} where a similar scaling function approaches a constant value, though in a singular fashion, at small argument.\\

Our numerical results show that the SCS occurs over a wide range of parameter values in the advection problem, and the power law which characterises the divergence is universal. It should be noted that our studies incorporate the important time-dependent correlations of the driving field. Moreover we show analytically that in the limit of a stationary interface, the equilibrium state of the sliding particles is an SCS, and determine the form of the scaling function. Remarkably, we find that the calculated equilibrium scaling function describes very well the SCS for the strongly nonequilibrium advection problem as well.\\

The evolution of the one-dimensional interface is described by the KPZ equation \cite{kpz}  
\begin{eqnarray}
{\partial h \over \partial t} = \nu {\partial^{2} h \over \partial x^{2}}  + {\lambda \over 2} 
({\partial h \over\partial x})^2 + \zeta_h(x,t) 
\label{kpz1}
\end{eqnarray}
where $h$ is the height field, $\zeta_h$ is a Gaussian white noise satisfying $\langle \zeta_h (x,t) \zeta_h({x}',t')\rangle = 2D_h \delta(x - {x}')\delta(t - t')$. If the $m^{th}$ particle is at position $x_m$, its motion is governed by  
 \begin{equation} 
  {dx_{m} \over dt} = \left.-a{\partial h \over \partial x } \right|_{x_{m}}+ \zeta_m (t)
  \label{passive} 
  \end{equation} 
  where the white noise $\zeta_m (t)$ represents the randomising effect
  of temperature, and satisfies $\langle \zeta_m (t) \zeta_m(t')\rangle
  = 2\kappa \delta(t - t')$. Equation~(\ref{passive}) is a strongly overdamped Langevin equation of a particle in a potential $h(x,t)$ that is also fluctuating, with $a$ determining the speed of sliding. In the limit when $h(x,t)=h(x)$ is static, a set of noninteracting particles would reach the equilibrium Boltzmann state with particle density $\sim exp(-\beta h(x))$ at late times where $\beta=a/\kappa$. On the other hand, when $h(x,t)$ is time dependent, the system eventually settles into a strongly nonequilibrium steady state. The transformation $v = -\partial h / \partial x$ maps Eq.~(\ref{kpz1}) (with $\lambda = 1$) to the Burgers equation, which describes a compressible fluid with local velocity $v$. The transformed Eq.~(\ref{passive}) describes passive scalar particles advected by the Burgers fluid. The ratio $a/ \lambda > 0$ corresponds to advection, the case of primary interest in this paper, while $a/ \lambda < 0$ corresponds to anti-advection (particles moving against the flow).\\

Rather than analysing the coupled Eqs.~(\ref{kpz1}) and ~(\ref{passive}) directly, we study a lattice model  which is expected to have similar behaviour at large length and time scales. The model consists of a flexible one-dimensional lattice in which particles reside on sites, while the links or bonds between
successive lattice sites are also dynamical variables which denote local slopes of the surface. The total number of sites is $L$. Each link takes values $+1$ (upward slope $\rightarrow /$) and $-1$ (downward slope $\rightarrow \backslash$). The rules for surface evolution are : for advection, choose a site at random, and if it is on a local hill $(\rightarrow /\backslash)$, change the local hill to a local valley$(\rightarrow \backslash /)$; otherwise leave it unchanged. For anti-advection, if a site is at a local valley, change it to a local hill; otherwise leave it unchanged ~\cite{drosselcomment2}. We use periodic boundary conditions, implying no overall tilt of the surface. After every $N_s$ surface moves we perform $N_p$ particle updates according to the following rule : we choose a particle at random and move it one step downward with probability $(1+K)/2$ or upward with probability $(1-K)/2$.  The parameter $K$ ranges from 1 (particles totally following the surface slope) to 0 (particles moving independently of the surface). We define the ratio $N_s/N_p$ as $\omega$. The continuum Eq.~(\ref{kpz1}), valid for $\omega=1$, gets modified for $\omega \neq 1$. While the first two terms on the R.H.S. of Eq.~(\ref{kpz1}) get rescaled by $\omega$, the noise term gets multiplied by $\sqrt{\omega}$ ~\cite{future}. The limit $\omega \rightarrow 0$ corresponds to the adiabatic limit of the problem where particles move on a static surface and the steady state is the thermal equilibrium state. In our simulations, we update the surface and particles at independent sites, reflecting the independence of the noises $\zeta_h (x,t)$ and $\zeta_m (t)$ ~\cite{drosselcomment}.\\

\begin{figure}
  \centering
  \includegraphics[width=0.7\columnwidth,angle=-90]{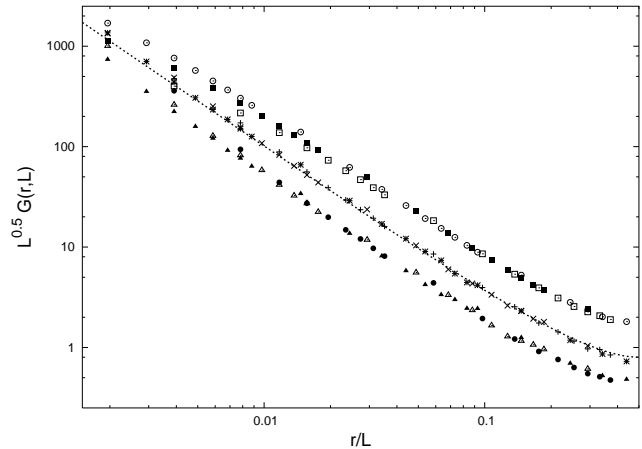}
  \caption{Two point scaled density correlation $G(r,L)$ function (advection) for $K=1$, $\omega=2$ (top curve), $1$ (middle), $1/2$ (bottom). The line is a plot of Eq.~(\ref{corrscaling}) with $\beta = 4$. The lattice sizes are $L~=~~256$ ($\Box$, $+$, $\bullet$), $512$ ($\blacksquare$, $\times$, $\triangle$) and $1024$ ($\circ$, $\ast$, $\blacktriangle$).
}
  \label{advcorr}

\end{figure}

 If $\omega$ is nonzero, the RMS displacement of each particle is proportional to $t^{1/z}$ where $z=3/2$ is the dynamical exponent of the KPZ surface ~\cite{chin, drossel2}. At large enough times $t \gg L^{z}$, the system settles into a nonequilibrium steady state and develops interesting correlations, to which we now turn. We begin by discussing results for advection, for $\omega = K = 1$. Figure~\ref{advcorr} shows the two-point (unconnected, scaled) density-density correlation function $G(r,L) = \langle n_i n_{i+r}\rangle_L$ where $n_i$ is the number of particles at site $i$. The total number of particles $N \equiv \sum n_{i}$ is taken to be equal to $L$. The numerical results give strong evidence that the scaling form 
\begin{eqnarray} 
G(r,L) \sim L^{-\theta} Y(r/L) 
\label{correlation}
\end{eqnarray}    
is valid for $r/L>0$ with $\theta \simeq {1/2}$. An unusual point is that the scaling function $Y(y)$ has a power law divergence $Y(y) \sim y^{-\nu}$ as $y \rightarrow 0$, with $\nu$ close to 3/2. Using this result, we can find the probability $p(r)$ of two particles being at a distance $r$ from each other in the limit $L \rightarrow \infty$, $p(r)=\frac{L}{N^{2}}G(r,L) \sim \frac{1}{r^{3/2}}$. This is in agreement with an exact result for two second class particles in the asymmetric exclusion process \cite{derrida}. Another related quantity is defined by first dividing the system into $L/l$ bins of of size $l$; then a particle is chosen at random and the number $N(l,L)$ of the particles lying in the same bin as this particle is measured~\cite{drossel2}. Using Eq.(\ref{correlation}), we find that $N(l,L) \sim c_{1}L(1-c_2l^{-\nu+1})$, in better agreement with the numerical results for $N(l,L)$ than the l-independent form of ~\cite{drossel2}.\\ 

\begin{figure}
  \centering
  \includegraphics[width=0.7\columnwidth,angle=-90]{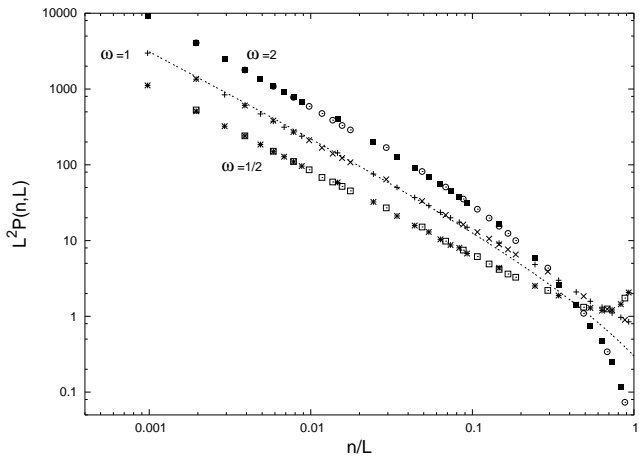}
  \caption{Scaled probability distribution $P(n,L)$ for $\omega = 1/2,1,2$ $(K=1)$. The line is a fit to Eq.~(\ref{gy1}) with $\beta=2.3$. The lattice sizes are $L$$=$ $512$ ($\circ$, $\times$, $\Box$), $1024$ ($\blacksquare$, $+$, $\ast$).
}
  \label{advdensity}
\end{figure}

\begin{figure}
  \centering
  \includegraphics[width=0.7\columnwidth,angle=-90]{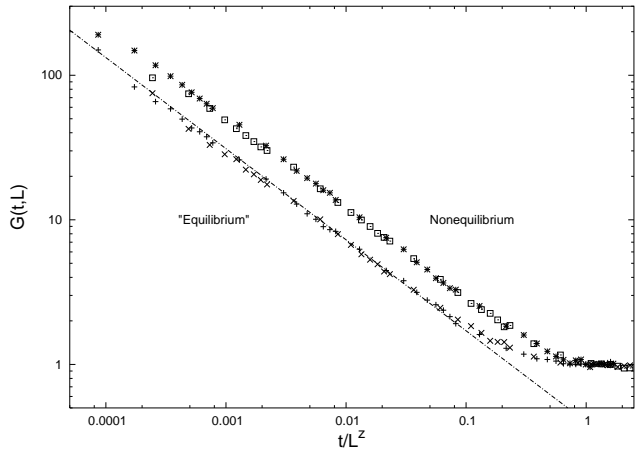}
  \caption{Scaled density-density autocorrelation function $ \widetilde G(t,L)$ (advection) in the nonequilibrium and equilibrium (adiabatic limit) cases. The dashed line shows $y \sim x^{-2/3}$. The lattice sizes are $L$$=$$256$ ($\Box$, $\times$), $512$ ($\ast$, $+$).
}
  \label{advauto}
\end{figure}

The central curve in Fig.~\ref{advdensity} shows the scaled probability  $P(n,L)$ that any site has occupancy $n$. This quantity, for $n>0$, shows the scaling form
\begin{eqnarray}
P(n,L) \sim {1\over L^{2 \delta}} f \left({n\over L^{\delta}}\right) 
\label{probability}
\end{eqnarray}
with $\delta =1$. As $y \rightarrow 0$, the scaling function $f(y)$ seems to behave as $y^{- \gamma}$ with $\gamma \simeq 1.15$. We will see below that the values of exponents and the functional forms of the scaling functions in Eqs.~(\ref{correlation}) and ~(\ref{probability}) agree surprisingly well with those for the equilibrium case. Note that Eq.~(\ref{probability}) leads to $\langle n^2 \rangle \equiv G(0,L)\sim L$ which is verified by direct simulation. From simulations we also find that the number of occupied sites $N_{occ} \equiv (1-P(0,L))L$ varies as $L^{\phi}$ with $\phi \simeq 0.23$, though the effective exponent seems to decrease systematically with increasing $L$. A scaling analysis yields $\delta=\nu-\theta$ and $\phi=\delta(\gamma-2)+1$ with the latter holding for $1 \geq \delta >0$ and $\gamma>1$. We also monitored the time dependent correlation function in the steady state. We find that (see Fig.~\ref{advauto}) the autocorrelation function $\widetilde{G}(t,L) \equiv \langle n_i(0) n_{i}(t)\rangle_{L}$ scales with system size as 
\begin{eqnarray} 
 \widetilde{G}(t,L) \sim \widetilde{Y} \left(t \over L^{z}\right).
\label{autocorrelation}
\end{eqnarray}  
with $z = 3/2$. The scaling function shows a power law behaviour $ \widetilde{Y}(\tilde y)\sim \tilde y^{- \psi}$ with $\psi \simeq 2/3$ as $\tilde y \rightarrow 0$.\\ 

These results lead to a simple picture of a typical SCS. The scaling of the probability distribution $P(n,L)$ with $n/L$ and the vanishing of the probability of finding an occupied site ($\equiv N_{occ}/L$) suggest that a large number of particles (often of the order of system size) accumulate on a few sites; the scaling of the two-point density-density correlation function with $L$ implies that the particles are distributed over distances of the order of $L$, while the divergence of the scaling function indicates clustering of large-mass aggregates. \\ 

Now let us turn to the limiting adiabatic case $\omega \rightarrow 0$ corresponding to an equilibrium system of particles at inverse temperature $\beta \propto \ln((1+K)/(1-K))$ distributed on a disordered, stationary surface. Instead of considering the number density of many noninteracting particles, we may consider equivalently the probability density of a single particle moving on the surface and average over all surface configurations, as in the Sinai model ~\cite{sinai}. For the KPZ equation in one dimension, the distribution of heights in the stationary state is described by ${\rm Prob} [\{h(r)\}]\propto \exp\left[-{\nu \over {2D_h}}\int \left(\frac{dh(r')}{dr'}\right)^2 dr'\right]$. Thus, any stationary configuration can be thought of as the trace of a random walker in space evolving via the equation, $dh(r)/dr = \xi(r)$ where the white noise $\xi(r)$ has zero mean and is delta correlated, $\langle \xi(r)\xi(r')\rangle = \delta (r-r')$. This is exactly the surface considered in the Sinai model. The probability $\rho (r) \equiv n_{r}/L$ of finding the particle at position $r$ is given by   $\rho(r)= \exp[-\beta h(r)]/Z$ with the partition function $Z=\int_0^L \exp[-\beta h(r')]dr'$.\\

The correlation function $ G(r,L)/L^2 = \langle \rho(r_0)\rho(r+r_0) \rangle$ involves an average over the surface configurations sampled from the stationary measure mentioned above:
\begin{eqnarray}
L^{-2} G(r,L)= \langle \left[{ {e^{-\beta [h(r_0)+h(r_0+r)]}}\over {Z^2}}\right] \rangle.
\label{eqcorr}
\end{eqnarray}
The right hand side of Eq.~(\ref{eqcorr}) was evaluated exactly by Comtet and Texier~\cite{comtet} in the context of one dimensional disordered supersymmetric quantum mechanics, where the right hand side of Eq.~(\ref{eqcorr}) is the correlation function in the ground state wave function. In the scaling limit, $r\to \infty$, $L\to \infty$ with the ratio $y=r/L$ fixed, one finds $G(r,L)\sim L^{-1/2} Y(r/L)$ where the scaling function is given by
\begin{eqnarray}
Y(y) = {1\over {\beta \sqrt{2\pi}}} [y(1-y)]^{-3/2}.
\label{corrscaling}
\end{eqnarray}
Note that the point $r=0$ is not part of the scaling function; we have $G(0,L)\approx \beta^2 L/{12}$. This formalism can also be used to calculate the equilibrium probability density $P(\rho,L)$ ~\cite{future}. Our results indicate that $P(\rho,L)$ can be written as the sum of two parts :
\begin{equation}
P(\rho,L) \approx \left[1- {{\ln^2 (L)}\over {\beta^2 L}}\right]\delta(\rho) + 
{4\over {\beta^4 L}} g\left[ {{2\rho}\over {\beta^2}}\right]\theta\left(\rho-{c\over {L}}\right),
\label{probscaling}
\end{equation}
The first part refers to vacant stretches, and to the fact that the number of occupied sites occupies a vanishing fraction $\sim (\ln L)^2/L$ of the system. The scaling function $g(y)$ in the second part is given by
\begin{equation}
g(y) = { {e^{-y}}\over {y} } K_0(y).
\label{gy1}
\end{equation}
where $K_0(y)$ is the modified Bessel function which has the asymptotic behaviour $[-\ln (y/2)-0.5772...]$ as $y \rightarrow 0$. The theta function incorporates a lower cutoff on the validity of the scaling form and $c$ is a constant of $O(1)$. Thus we see that in the $\omega \rightarrow 0$ limit, the equilibrium state is an SCS.\\ 

Surprisingly, these equilibrium results reproduce quite well the scaling exponents and scaling functions for $G(r)$ and $P(n)$ for $n \geq 1$ obtained numerically for the nonequilibrium case $\omega = K = 1$, as can be seen in Figs.~\ref{advcorr} and ~\ref{advdensity}, though with different values of $\beta$. The correlation function matches with $\beta \simeq 4$ while $\beta \simeq 2.3$ describes the probability distribution of number data well. However, $P(0,L)$ (and thus $N_{occ}$) does not agree closely in the two cases. The equilibrium case can also be used to shed light on the dynamical properties of the nonequilibrium steady state. We compared our results for $G(t,L)$ with the density-density autocorrelation function in the adiabatic $\omega \rightarrow 0$ limit. To find the latter, we simulated a surface with height field $h(r,t)$ evolving according to KPZ dynamics, and evaluated the density using the equilibrium weight $\rho(r,t)= \exp[-\beta h(r,t)]/Z$. As shown in Fig.~\ref{advauto}, the results with $\beta = 4$ agree with the autocorrelation function in the nonequilibrium system, apart from a numerical factor.\\

Why do the equilibrium results describe the non-equilibrium steady state so well? A partial explanation may lie in the fact that the statistical properties of the valleys where the particles settle down in both the cases are similar; under advection dynamics, particles slide down and reach the valleys relatively quickly, allowing them to explore the terrain. So, at least in a certain range of parameter space, the surface fluctuation noise mimics the effect of temperature insofar as it causes redistribution of particles over the landscape. Intriguingly, the temperature depends on the property under consideration --- a phenomenon that deserves further study.\\

The close correspondence between the SCS obtained in the nonequilibrium case and the equilibrium case $\omega=0$ suggests there may be some universal features as parameters are varied. We confirm this by simulations, varying $\omega$ in the regime $1/4 \leq \omega \leq 4$. We find that for large $L$ scaling is valid, with scaling functions which have universal exponents characterising singular behaviour at small values of $r/L$ (for $G(r,L)$) and $n/L$ (for $P(n,L)$). Nonuniversal features emerge at large values of the arguments $r/L$ and  $n/L$. For $\omega<1$, we find that $P(n)$ is nonmonotonic and shows a peak at large $n$ (Fig.~\ref{advdensity}). Numerically, we find that $P(n=N)/P(n=N-1) = 1 / \omega$ (for $\omega \leq 1$), which can be argued for by considering the processes which lead to the formation of an $N$ particle cluster from an $N-1$ particle cluster and $\it{vice}$ $\it{versa}$ ~\cite{future}.\\ 

We have also studied anti-advection, where particles move opposite to the direction of surface motion. We again find that the density-density correlation function follows the scaling form of Eq.~(\ref{correlation}) but with $\theta = 0$. The scaling function $Y(r/L)$ diverges as $(r/L)^{-\nu}$ with $\nu \simeq 0.33$. Thus the steady state for anti-advection is also an SCS; the correlation function depends strongly on the size, and does not follow a size independent power law as in ~\cite{drossel1}. It is also interesting to study passive sliders on an Edwards-Wilkinson surface ~\cite{manoj} instead of the KPZ surface. Our simulations on this problem show an SCS having the scaling form of Eq.~(\ref{correlation}) with $\theta = 0$, and $\nu \simeq 0.66$. As for advection, the relation $\delta =\nu-\theta$ holds in both these cases. \\

In conclusion, we remark on two open problems. First, it would be worthwhile to understand better the correspondence we have noted between the SCS in the nonequilibrium advection and the SCS in the equilibrium, Sinai problem, and in particular why different quantities in the former involve different temperatures in the latter. Secondly, the existence and extent of clustering induced by fluctuating potentials in higher dimensions remain to be studied and established.


\begin{thebibliography}{40}
\bibitem[{\citenamefont{Drossel and Kardar}(2000)}]{drosselcomment}
 This is in contrast to~\cite{drossel2} where only particles at a site affected by the surface evolution are moved.
\end{thebibliography}
\end{document}